\documentclass[a4paper,fleqn]{cas-sc}

\usepackage[authoryear,longnamesfirst]{natbib}
\usepackage[american]{babel}
\usepackage{setspace}
\usepackage{amsthm,amsmath,amsfonts,amssymb}
\usepackage{graphicx}
\usepackage{multirow}
\usepackage{subcaption}
\newtheorem{theorem}{Theorem}
\newtheorem{lemma}[theorem]{Lemma}

\newcommand{\DEP}{{\rm D_{EP}}( f\parallel g)}
\newcommand{\dep}{{\rm D_{EP}}}
\newcommand{\PED}{{\rm D_{EP}}( g\parallel f)}
\def\tsc#1{\csdef{#1}{\textsc{\lowercase{#1}}\xspace}}
\tsc{WGM}
\tsc{QE}
\tsc{EP}
\tsc{PMS}
\tsc{BEC}
\tsc{DE}

\begin{document}
\let\WriteBookmarks\relax
\def\floatpagepagefraction{1}
\def\textpagefraction{.001}

\shorttitle{}

\shortauthors{Datta }

\title [mode = title]{Bayes Factor Functions for Testing Partial Correlation Coefficients}                      



%
\author[]{Saptati Datta}[
                        orcid=0009-0009-3331-6523]

\cormark[1]


\ead{saptati@tamu.edu}


\credit{ Methodology, Implementation}


\affiliation[]{organization={Texas A\&M University}, 
    city={College Station},
    postcode={3143 TAMU}, 
    country={USA}}




\cortext[cor1]{Corresponding author}



\begin{abstract}
Partial correlations are widely used in psychology and related fields to evaluate the relationship between two variables while controlling for others. In this article, we extend Bayes Factor Functions (BFFs; Johnson, Pramanik, \& Shudde) to the assessment of partial correlation. BFFs provide Bayes factors derived from test statistics and expressed as functions of standardized effect sizes. Unlike $p$-values, which have been criticized for failing to accumulate evidence for the true hypothesis, and unlike conventional Bayesian methods, which can be computationally demanding and sensitive to prior specification, BFFs offer a practical objective alternative. They summarize evidence across a range of prior distributions on standardized effects and allow researchers to integrate results across studies. We further emphasize that the proposed test for partial correlations differs from conventional ($t/F$) tests, since the distribution of the test statistic under the alternative hypothesis follows a nonstandard form. We demonstrate how BFFs can be applied to partial correlation testing to provide interpretable and cumulative evidence in psychological research.

\end{abstract}




\maketitle

\section{Introduction}

Partial correlations are frequently employed in social science and psychological research to quantify the association between two variables while statistically controlling for the influence of others. {This concept is also fundamental in Gaussian graphical models\citep{lauritzen1996, Whittaker1990GraphicalModels}, where one tests for the presence of an association between two variables while conditioning on all others. \cite{Epskamp2017} introduce a network model in psychometrics for multivariate normal data, conceptualizing the covariance between psychometric indicators as arising from pairwise interactions between observable variables in a network structure, in contrast to standard psychometric models where covariance between test items arises from the influence of one or more common latent variables; \cite{EpskampGGM2018} introduce Gaussian graphical models (GGMs) as undirected networks of partial correlation coefficients and show their applicability to cross-sectional, time-series, and multilevel data; \cite{Borsboom2021NetworkAnalysis} review network models (including GGMs) in psychology and discuss applications in personality, mental health, and attitudes, arguing that they can be used to explore the structure of high-dimensional data in the absence of strong prior theory on how variables are related, complementing exploratory techniques such as exploratory factor analysis (representing shared variance due to a small number of latent variables) and multidimensional scaling (representing similarity relations between objects in a low-dimensional metric space), while uniquely focusing on patterns of pairwise conditional dependencies, providing effective visualizations of multivariate statistical associations, and generating causal hypotheses by linking conditional independence relations to potential causal dynamics.} Existing approaches to testing partial correlations typically rely on regression-based methods. One common strategy involves computing a test statistic and defining a Bayes factor based on Cohen's $f^2$ \citep{KlauerMeyerGrantKellen2025}. {Several classical approaches exist for testing whether a partial correlation coefficient is zero without relying on regression formulations. One common method is based on the Fisher $z$-transformation of the sample partial correlation,  which yields an asymptotically normal test statistic under the null hypothesis. Likelihood-based procedures arising from the multivariate normal model provide another route, where testing a zero partial correlation is equivalent to testing whether the corresponding entry of the precision matrix is zero, leading to likelihood ratio, Wald, or score tests. Finally, exact finite-sample tests can be derived from the Wishart distribution of the sample covariance matrix, allowing inference on entries of the precision matrix without invoking regression representations. We refer the reader to \cite{Anderson} for detailed discussions of these methods.} All of these approaches yield $p$ values that serve as the basis for inference. However, $p$ values provide no direct measure of evidence in support of the null hypothesis of no partial correlation—an inference that is often of substantive importance in psychological research\citep{Rouder2009, MoreyRouder2011, wetzels2012, Wagenmakers2016}. {
Bayes factors \citep{KassRaftery1995, Johnson2005, Johnson2023}, on the other hand, represent the ratio of the marginal probability assigned to the observed data by two competing hypotheses and, when combined with the prior odds assigned between the hypotheses, yield the posterior odds that each hypothesis is true. Specifically, if $P(H_0)$ and $P(H_1)$ denote the prior probabilities of $H_0$ and $H_1$, respectively, then
\begin{align}\label{eq: BF-postodds}
\frac{P(H_1 \mid \mathbf{y})}{P(H_0 \mid \mathbf{y})}
=
\frac{m_1(\mathbf{y})}{m_0(\mathbf{y})}
\;
\frac{P(H_1)}{P(H_0)} .
\end{align}
Equivalently,
\begin{align}
\text{Posterior Odds}
=
BF_{10}(\mathbf{y}) \times \text{Prior Odds},
\end{align}
where the Bayes factor is defined as $BF_{10}(\mathbf{y}) = m_1(\mathbf{y})/m_0(\mathbf{y})$, with $m_0(\mathbf{y})$ and $m_1(\mathbf{y})$ denoting the marginal probabilities assigned to the data under $H_0$ and $H_1$, respectively. The corresponding posterior probability of $H_1$ is $P(H_1 \mid \mathbf{y}) = \dfrac{BF_{10}(\mathbf{y}) P(H_1)}{P(H_0) + BF_{10}(\mathbf{y}) P(H_1)}$, and $P(H_0 \mid \mathbf{y}) = 1 - P(H_1 \mid \mathbf{y})$.}



To overcome the shortcomings of $p$ values, several Bayesian tests for the presence of partial correlations have been developed. 
These include, among others, \cite{wetzels2012} and \cite{Wang2019}, which are based on a correlation model that can be described as follows:
\begin{equation}\label{model}
    Y = \alpha + \beta_1 X_1 + \beta_2 X_2 + \epsilon,
\end{equation}
where $\alpha$ represents the intercept, $\epsilon$ is a normally distributed error with precision $\psi$, and the predictors $X_1$ and $X_2$ are centered.  {\cite{Wang2019} impose an additional orthogonality condition \(X_1^\top X_2 = 0\), under which the alternative prior in \eqref{eq:prior_sp_liang} is replaced by the specification \(\beta_2 \mid \psi, g \sim \mathcal{N}\!\big(0,\;\tfrac{g}{\psi}(X_2^\top X_2)^{-1}\big)\). Here, the scalar $g$ serves as a variance scaling parameter that regulates the degree of prior uncertainty assigned to the regression coefficients relative to the information in the data. In the context of hypothesis testing and model comparison, $g$ governs the allocation of prior mass over effect sizes. Smaller values of $g$ concentrate mass near zero, yielding behavior akin to a local prior, whereas larger values of $g$ diffuse the prior over a broader range of parameter values, resulting in a slower accumulation of evidence in favor of the true hypothesis.} In this framework, testing for a partial correlation amounts to evaluating whether $\beta_2$ differs from zero once the effect of $X_1$ has been accounted for. This can be expressed as a model selection problem between

\begin{align}
    M_0:\; Y=\alpha+\beta_1X_1+\epsilon,\qquad \epsilon\sim\mathcal{N}(0,\psi^{-1}I_n), \nonumber \\
M_1:\; Y=\alpha+\beta_1X_1+\beta_2X_2+\epsilon,\qquad \epsilon\sim\mathcal{N}(0,\psi^{-1}I_n).
\end{align}

Under both models, the conventional prior on $(\alpha,\psi)$ is given by $p(\alpha,\psi)\propto \psi^{-1}$, corresponding to a flat prior on $\alpha$ and Jeffreys’s prior for $\psi$.  In most of the existing literature, the regression coefficients are assigned Zellner’s $g$-prior \citep{Liang2008}. { The priors on the parameters under $M_0$ and $M_1$ are specified as: 
\begin{align}\label{eq:prior_sp_liang}
    &M_0:\pi\left(\beta_1, \psi, g  \right) \propto \frac{1}{\psi} \\
    &M_1: \begin{pmatrix}\beta_1\\ \beta_2\end{pmatrix}\Bigm|\psi,g \sim \mathcal{N}\!\Big(0,\;\tfrac{g}{\psi}(X^\top X)^{-1}\Big),
\end{align}
 where $X=(X_1,X_2)$.}    However, with a fixed $g$, two paradoxes arise \citep{Liang2008}. As $g\to\infty$, the prior on $\beta_2$ becomes arbitrarily diffuse, and the Bayes factor collapses to zero, automatically favoring $M_0$ regardless of the data (Bartlett’s paradox). {For inference within a fixed model, the posterior distribution may remain reasonable even when \(g\) is taken to be very large in an attempt to represent a noninformative prior. In the context of model selection, however, such a choice is generally undesirable. In particular, in the limiting case where \(g \to \infty\) while \(n\) and \(p\)(dimension of the parameter space under $M_1$) remain fixed, the Bayes factor comparing \(M_{1}\) with \(M_{0}\) converges to \(0\). Consequently, the excessive dispersion of the prior induced by choosing a very large \(g\) unintentionally forces the Bayes factor to favor the null model—the smallest model—regardless of the information contained in the data. This phenomenon was noted by \cite{Bartlett1957} and is commonly referred to as \emph{Bartlett's paradox}, which was further discussed by \cite{Jeffreys1961}.} Conversely, when the data overwhelmingly support $M_1$ so that $R_1^2\to 1$, the Bayes factor converges to a constant instead of diverging to infinity (information paradox). To avoid these pathologies, \cite{wetzels2012} applied the Jeffreys–Zellner–Siow(JZS) prior, proposed by \cite{Liang2008}, which represents a Cauchy distribution on regression coefficients as a mixture of $g$-priors with $
g \sim \mathrm{Inv\text{-}Gamma}\!\left(\tfrac{1}{2},\tfrac{n}{2}\right).
$ Integrating out $g$ yields the JZS Bayes factor for partial correlations, which preserves tractability while ensuring that $BF_{10}\to\infty$ as $R_1^2\to 1$ and avoids automatic preference for the null when $g$ is large, thus resolving both paradoxes. With this specification, the Bayes factor can be written as

$$
BF_{10} \;=\; 
\frac{\displaystyle\int_0^\infty (1+g)^{\tfrac{n-1-p_1}{2}}\,[1+(1-R_1^2)g]^{-\tfrac{n-1}{2}}\,g^{-3/2}e^{-n/(2g)}\,dg}
{\displaystyle\int_0^\infty (1+g)^{\tfrac{n-1-p_0}{2}}\,[1+(1-R_0^2)g]^{-\tfrac{n-1}{2}}\,g^{-3/2}e^{-n/(2g)}\,dg},
$$

where $R_0^2$ and $R_1^2$ are the coefficients of determination under $M_0$ and $M_1$, and $p_0,p_1$ are the corresponding numbers of regression coefficients. Here, $R_0^2=r_{YX_1}^2$ and $
R_1^2 \;=\; r_{YX_1}^2 \;+\; \frac{r_{YX_2\mid X_1}^2}{\,1-r_{YX_1}^2\,},$ where $r_{YX_1}^2$ is the sample correlation coefficient between $Y$ and $X_1$, and $r_{YX_2\mid X_1}^2$ is sample partial correlation coefficient between $Y$ and $X_2$ keeping $X_1$ fixed. However, this approach produces results that depend on the choice of predictor variable, making the outcome sensitive to the direction of the effect. This dependence is undesirable, since partial correlation is inherently an undirected measure of association. { For any collection of random variables $(Y, X_1, X_2)$, irrespective of whether they arise from a linear regression model of the form \eqref{model}, the partial correlation quantifies the strength of association between two variables after adjusting for the linear effects of the remaining variables. Specifically, the partial correlation between $Y$ and $X_2$ given $X_1$ is defined as $\rho_{Y,X_2 \mid X_1} = \operatorname{Cov}(Y,X_2 \mid X_1)\big/\sqrt{\operatorname{Var}(Y\mid X_1)\operatorname{Var}(X_2\mid X_1)}$. This quantity is symmetric in the variables being compared, so that $\rho_{Y,X_2 \mid X_1} = \rho_{X_2,Y \mid X_1}$. Consequently, partial correlation quantifies conditional association rather than directional prediction: it measures the strength of dependence between two variables after adjusting for the effect of other variables without assigning one variable the role of outcome and the other the role of predictor. This symmetry is fundamental in multivariate analysis and underlies the interpretation of partial correlations in Gaussian graphical models, where conditional associations are represented by undirected edges between variables \citep{lauritzen1996,Epskamp2018,Borsboom2021NetworkAnalysis}. In contrast, regression-based formulations require one variable to be designated as the response, which introduces an artificial directionality. Quantities such as $R^2$ therefore depend on which variable is treated as the outcome in the regression model. When the Bayes factor for testing a partial correlation is expressed through such regression quantities, the resulting test may depend on this arbitrary choice of response variable. This dependence is undesirable because the scientific question concerns the presence of a conditional association between variables, which is inherently symmetric, rather than a directional predictive relationship.}

\cite{Kucharsky2023PartialCorrelation} proposed an analytic Bayesian framework for inference on Pearson partial correlations, providing closed-form expressions for both the Bayes factor and the marginal posterior of the population coefficient. The method employs a \emph{stretched-$\beta$ prior} \eqref{eqn: sbprior}, obtained by mapping a $\text{Beta}(\alpha,\alpha)$ distribution from $[0,1]$ to $[-1,1]$, thereby producing a symmetric prior centered at zero. The hyperparameter $\alpha$ controls the concentration of prior mass: when $\alpha=1$, the prior is uniform over $[-1,1]$, whereas larger $\alpha$ values increasingly concentrate probability near $\rho=0$. This choice allows analytic tractability, but it also renders the Bayes factor sensitive to the specification of $\alpha$. In particular, information consistency---the requirement that the Bayes factor diverges when $r_{xy.z}=\pm 1$---holds only when $\alpha \leq 1/2$, whereas for $\alpha>0$ divergence occurs only if $n-k \geq 2\alpha+2$. As a result, the strength of evidence provided by the Bayes factor can vary substantially with prior concentration, underscoring the importance of robustness checks when applying this methodology. { We note that the \emph{stretched beta prior} is a \emph{local prior} which assigns non-negligible mass to parameter values consistent with the null hypothesis under the alternative, and consequently, the Bayes factor against the null hypothesis exhibits a relatively slow convergence rate of order \(O_p(n^{-0.5})\) when the null is true \citep{Johnson2010, DATTA2025}. To address this limitation, we instead employ non-local priors \eqref{sec: NLP} for the parameter of interest, which ensure a faster rate of convergence under the null hypothesis. We will also demonstrate later in this manuscript that non-local priors facilitate expressing Bayes factors as functions of standardized effect size, since the mode of such priors can be determined based on the value of the standardized effect size. In contrast, the mode of a \emph{stretched beta prior} is fixed at the null value of 0. }

Furthermore, Bayes factors derived from the full data likelihood necessitate the specification of priors on nuisance parameters; such choices are inherently subjective, and differing prior specifications can lead to different values of the Bayes factor. This sensitivity is exacerbated as the number of nuisance parameters increases. To navigate this issue, we use Bayes factors based on test statistics \citep{Johnson2005} to {assess for the evidence} of partial correlation coefficients. This strategy bypasses the requirement to specify prior distributions on nuisance parameters. \citet{Johnson2005} introduced this framework for constructing Bayes factors based on commonly used test statistics by assigning priors to the noncentrality parameters that characterize the corresponding alternative hypotheses. In this approach, Bayes factors are formulated using standard \( z \)-, \( t \)-, \( \chi^2 \)-, and \( F \)-statistics. While the sampling distributions of these statistics under the null hypothesis are well known, their asymptotic distributions under the alternative hypotheses are determined by an unknown non-centrality parameter. {A \emph{non-centrality parameter} is a parameter that indexes the sampling distribution of a test statistic under the alternative hypothesis and quantifies the magnitude of departure from the null value. For example, if \(Z\) is a \(z\)-statistic, then under \(H_0\), \(Z \sim N(0,1)\), whereas under the alternative \(Z \sim N(\delta,1)\), where \(\delta\) is the non-centrality parameter. Similarly, a \(t\)-statistic follows a central \(t_\nu\) distribution under the null hypothesis, but under the alternative it follows a non-central \(t_\nu(\delta)\) distribution indexed by the non-centrality parameter \(\delta\).}  Consequently, specifying a prior for the alternative hypothesis becomes straightforward, and no prior is required under the null hypothesis.

{To test for the  of partial correlations, \cite{KlauerMeyerGrantKellen2025} construct a Bayes factor based on the classical partial \(F\)-test for comparing reduced and full regression models, where the test statistic follows a central \(F\) distribution under \(H_0\) and a non-central \(F\) distribution under \(H_1\). The non-centrality parameter under the alternative is expressed in terms of Cohen’s \(f^2\). The Bayes factor is then obtained by integrating the non-central \(F\) density with respect to a prior on this non-centrality parameter. Similar to \cite{Wagenmakers2016, Wang2019}, this construction is too therefore inherently regression-based, relying on residuals from nested models and requiring specification of a response variable and predictors, rather than yielding a symmetric test defined through the joint distribution of multivariate observations. Although the substantial body of related work underscores the importance of developing such methodologies, there remains a clear need for approaches that directly test whether a given pair of variables is conditionally independent given the rest when the data consist solely of multivariate observations \((\mathbf{Z_1},\dots, \mathbf{Z_p}) \sim \mathcal{N}_p(0,\boldsymbol{\Sigma})\). For instance, in a GGM, one  encodes conditional independence structure through \(\boldsymbol{\Theta}=\boldsymbol{\Sigma^{-1}}\), where an edge between nodes \(j\) and \(k\) is present if and only if \(\Theta_{jk}\neq 0\). The inferential task is to determine which entries of \(\Theta\) are zero, which is inherently a statement about the joint distribution of all variables and does not involve selecting a response variable or assessing how much variance in one variable is explained by others. In contrast, a regression-based \(F\)-test evaluates whether predictors improve prediction of a designated response, whereas a GGM tests whether two variables are conditionally independent given all others; these constitute fundamentally different statistical objects. }

We propose a residual-free method to construct Bayes factors using a sufficient statistic for partial correlations. In our formulation, the test statistic \eqref{eq:t_def}, which is a function of the sample partial correlation coefficient, does not follow a non-central $t$ distribution under the alternative hypothesis \cite{Anderson}. In Section 2.3, we note that since the sampling distribution of the test statistics used to evaluate the significance of partial correlations deviates from standard forms under the alternative hypothesis, it necessitates the introduction of a noncentrality parameter.

{ A well-recognized criticism of Bayes factors is that their construction typically requires specification of a simple alternative hypothesis in order to induce an informative/subjective prior. We address this limitation by formulating Bayes factors as functions of effect sizes evaluated over a continuum of alternative specifications \citep{Johnson2023}.} To study the plausibility of a range of  alternative hypotheses, we calculate Bayes factor functions (BFFs) to illustrate the evidence provided by a range of alternative prior distributions centered on standardized effect sizes of interest \citep{Johnson2023}. This methodology shares commonalities with that proposed in \citep{Franck2020}, where Bayes factors are expressed as functions of hyperparameters.  The primary differences between the approaches are that BFFs avoids prior specifications on nuisance parameters by modeling test statistics directly and by imposing prior distributions centered on effect sizes, which are often the primary parameters of interest in null hypothesis significance tests.  Following the BFF prescription, we express Bayes factors as a function of effect sizes by equating the modes of prior distributions to a function of standardized effect sizes.

Our primary contribution is the formulation of Bayes factor functions based on test statistics, recognizing that testing partial correlations differs from conventional \( t \)- or \( F \)-tests due to the nonstandard sampling distribution of the test statistic under the alternative hypothesis. We now briefly review the statistical concepts used in the development of our methodology.

\subsection{Non-local priors}\label{sec: NLP}

 Non-local alternative prior (NAP) densities are essential for rapidly gathering evidence that supports either the true null or the true alternative hypothesis (\cite{Johnson2010},\cite{Rossell2017}) and defining Bayes factor functions. These densities are zero when the noncentrality parameter of a test statistic is zero. This permits faster accumulation of evidence supporting both true null and true alternative hypotheses. The characteristics of these densities are examined in \cite{Johnson2010}. Specifically, \cite{Johnson2010} delves into two varieties of non-local priors: \emph{moment prior} and
 \emph{inverse moment prior densities}. We will use normal moment priors to define Bayes factor functions for partial correlations.   It is defined below and is denoted by $\pi_{nm}$.

 \emph{Moment Prior Densities:} Let $\theta$ denote the parameter of interest and $\Theta$ be the parameter space. Let $\pi_b(\theta)$ be the base density with two bounded derivatives in a neighborhood containing $\pi_b(\theta_0)$, where $\theta_0$ is the value of the parameter consistent with the null hypothesis and $\pi_b(\theta_0)>0$. Then the $\nu^{th}$ order moment prior density is defined as
 \begin{eqnarray}\label{eqn:mom_prior}
     \pi_{nm}(\theta \,|\, \theta_0, \tau_\nu, \nu) = \frac{|\theta - \theta_0|^{2\nu}}{\tau_\nu}\pi_b(\theta),
 \end{eqnarray}
 where $\tau_\nu = \int_\Theta |\theta - \theta_0|^{2\nu}\pi_b(\theta) \ d\theta $. When $\theta_0 = 0$,  for $\nu \geq 1, \tau_\nu^2 > 0$, the  normal prior prior density \citep{DATTA2025} is defined as
\begin{equation}\label{nm_prior}
    \pi_{nm}(\theta \,|\, \tau^2, \nu) = \frac{|\theta|^{2\nu}}{\left(2\tau^2\right)^{\nu + 0.5}\Gamma\left(\nu + 0.5\right)} \exp\left(-\frac{\theta^2}{2\tau^2}\right)
\end{equation}

 When the alternative prior on $\theta$ is continuous and strictly positive at zero—corresponding to a continuous local alternative—and under certain regularity conditions, the Bayes factor in favor of the alternative hypothesis, when the null is true, converges at the rate $BF_{10} = O_p(n^{-1/2})$ \citep{Johnson2010}. For specific test statistics, such as the $z$, $t$, $\chi^2$, and $F$ statistics, \citet{DATTA2025} derived closed-form expressions for Bayes factors based on the $\nu$-th order normal moment priors
imposed on noncentrality parameters $\lambda$ of $z$ and $t$ tests, and $Gamma(\lambda\mid k/2+r,1/(2\tau^2)$ priors for $\chi^2_k$ and $F_{k,m}$ tests \citep{DATTA2025, chakraborty2025differentiallyprivatebayesiantests}.
Their results show that Bayes factors in favor of a true null hypothesis satisfies $BF_{10} = O_p(n^{-\nu - 1/2})$ for $z$ and $t$ tests, and $BF_{10} = O_p(n^{-\nu - k/2})$ for $\chi^2_k$ and $F_{k,m}$ tests. {A class of local priors that assigns positive mass at $0$, while permitting the mode to be specified in terms of a standardized effect size consistent with the alternative, is given by effect size priors \citep{KlauerMeyerGrantKellen2025}. This structure facilitates the formulation of Bayes factor functions, which require prior modes to be placed on parameter values aligned with the alternative hypothesis.
In Section \ref{sec:application}, we compare the Bayes factor functions obtained under the normal moment prior and the effect size prior of \cite{KlauerMeyerGrantKellen2025} in \eqref{eqn:esprior}, thereby illustrating the manifestation of these asymptotic rates.}




{
\subsection{Bayes factor based on Test Statistic}

Bayes factors based on test statistics (BFBOTS) \citep{Johnson2005} are Bayes factors defined in terms of standard test statistics such as the $z$, $t$, $\chi^2$, and $F$ statistics. The sampling distributions of these statistics under the null hypothesis are completely known. Under the alternative hypothesis, their distributions are typically non-central versions of the same families, characterized by a scalar non-centrality parameter. This representation simplifies prior specification under the alternative hypothesis because the prior can be placed directly on the non-centrality parameter, and no prior distribution is required under the null hypothesis.

To illustrate, suppose $X_1,\ldots,X_n \sim \mathrm{N}(\mu,\sigma^2)$ and consider testing $H_0:\mu=0$ against $H_1:\mu\neq0$. The statistic $t=\frac{\sqrt{n}\bar X}{\hat\sigma}$, where $\bar X$ and $\hat\sigma$ denote the sample mean and sample standard deviation, follows a Student's $t_{n-1}(0)$ distribution with $n-1$ degrees of freedom under $H_0$. Under $H_1$, the same statistic follows a non-central $t_{n-1}(\delta)$ distribution with non-centrality parameter $\delta \neq 0$. Thus the null hypothesis corresponds to $\delta=0$, whereas under the alternative $\delta>0$. Since the non-centrality parameter satisfies $\delta=\frac{\sqrt{n}\mu}{\sigma}$, testing $H_0:\mu=0$ against $H_1:\mu\neq0$ is equivalent to testing $H_0:\delta=0$ against $H_1:\delta\neq0$. Consequently, prior distributions can be specified directly on the non-centrality parameter $\delta$, allowing Bayesian testing to be conducted without introducing priors on nuisance parameters under the null. If $P(H_0)$ and $P(H_1)$ denote the prior probabilities of $H_0$ and $H_1$, respectively, then
\begin{align}
\frac{P(H_1 \mid \mathbf{t})}{P(H_0 \mid \mathbf{t})}
=
\frac{m_1(\mathbf{t})}{m_0(\mathbf{t})}
\frac{P(H_1)}{P(H_0)} .
\end{align}

Equivalently,
\begin{align}
\text{Posterior Odds}
=
BF_{10}(\mathbf{t}) \times \text{Prior Odds},
\end{align}
where the Bayes factor is defined as $BF_{10}(\mathbf{t}) = m_1(\mathbf{t})/m_0(\mathbf{t})$, with $m_0(\mathbf{t})$ and $m_1(\mathbf{t})$ denoting the marginal probabilities assigned to the test statistic under $H_0$ and $H_1$, respectively. The corresponding posterior probability of $H_1$ is
\begin{align}
P(H_1 \mid \mathbf{t})
=
\frac{BF_{10}(\mathbf{t}) P(H_1)}
{P(H_0) + BF_{10}(\mathbf{t}) P(H_1)},
\end{align}
and $P(H_0 \mid \mathbf{t}) = 1 - P(H_1 \mid \mathbf{t})$.


The only distinction between BFBOTS and Bayes factors derived from the full data therefore lies in how the posterior probability of the hypotheses are defined. BFBOTS quantify posterior probabilities of the hypotheses given the test statistic, whereas the conventional Bayes factor quantifies posterior probabilities given the full data. However, in BFBOTS the test statistics used are sufficient statistics that capture the entire information contained in the data. Consequently, the Bayes factor based on the test statistic retains the same inferential interpretation as the Bayes factor derived from the full data.
} 
\subsection{Bayes factor functions}\label{sec:BFF}
{ A commonly noted limitation of Bayesian hypothesis testing is that specifying an appropriate prior under the alternative hypothesis typically necessitates defining a fully specified alternative—i.e., identifying a particular alternative around which the prior is centered \citep{Johnson2023}—rather than working with a one- or two-sided composite alternative.  Bayes factor functions circumvent this difficulty by allowing the prior mode to vary across the parameter space under the alternative. To satisfy this requirement, \cite{Johnson2023} defined Bayes factor functions(BFFs), which express Bayes factors as a function of standardized effect size by centering prior densities  on a range of values of the non-centrality parameter allowed under the alternative hypothesis.} To illustrate the construction of a Bayes factor functions, consider a $t$ test as outlined by \cite{DATTA2025}:

Denote the probability distribution of a test statistic $t$ under the null and alternative hypotheses as follows, where $T_\nu(\lambda)$ denotes a T distribution on $\nu$ degrees-of-freedom and non-centrality parameter $\lambda$:
\begin{eqnarray}
H_0: t &\sim& T_\mu(0), \\
H_1: t | \lambda &\sim& T_\mu(\lambda), \quad \text{where} \quad \lambda | \tau^2 \sim \pi_{nm}(\lambda| \tau^2, \nu), \quad \text{with} \quad \tau>0, \quad r \geq 1.
\end{eqnarray}
Under these assumptions, it follows that the Bayes factor favoring the alternative hypothesis is given by
\begin{eqnarray}
    BF_{10}(t \mid \tau^2,r) &=& c  {_2}F_1\left(\frac{\mu+1}{2}, \nu+\frac{1}{2}, \frac{1}{2}, y^2\right) ,
\end{eqnarray}
where 
\begin{equation}
    y = \frac{\tau t}{\sqrt{(\nu+ t^2)(1+\tau^2)}} \quad \text{and} \quad c = \frac{1}{(1+\tau^2)^{r+\frac{1}{2}}}
\end{equation}
and ${_2}F_1(a,b,c; z)$ denotes the Gaussian hypergeometric function.

\cite{Johnson2023} suggest a choice of $\tau^2$ to ensure that the mode of the prior distribution for the non-centrality parameter aligns with a predetermined standardized effect size. To demonstrate, consider a $t$ test for testing the null hypothesis $H_0: \theta=0$, using a random sample $x_1,\dots,x_n$ from a normal distribution $N(\theta,\sigma^2)$, where $\sigma^2$ is not known. In this scenario, the test statistic $t$ is calculated as $t=\sqrt{n} \bar{x}/s$, with $s^2$ representing the commonly used unbiased estimator for $\sigma^2$. The distribution of $t$, given $\mu$ and $\sigma$, is
\[
t\mid\mu, \sigma \sim T_{\mu}\left(\frac{\sqrt{n}\theta}{\sigma}\right),
\]
where $\mu=n-1$ denotes the degrees of freedom. Under the null hypothesis, the distribution of $t$ follows $T_{\nu}(0)$. The non-centrality parameter for the $t$ distribution under the alternative hypothesis is denoted as $\lambda= \sqrt{n}\omega$, where $\omega=\theta/\sigma$ represents the standardized effect size. The Bayes factors are functions of these hyperparameters, and in the given example the Bayes factor \(BF_{10}(t \mid \tau^2,\nu)\) depends on \(\tau^2\), which in turn is a function of the standardized effect size, \(\omega\). { For comparison, we also employ the effect size prior of \cite{KlauerMeyerGrantKellen2025} in \eqref{eqn:esprior}, with its mode specified through standardized effect sizes, to construct Bayes factor functions over a sequence of alternatives.

BFFs therefore allow the calculation of a range of Bayes factors by varying the prior densities imposed on the noncentrality parameter used to define the alternative hypothesis. The family of prior densities used to define the Bayes factors is indexed by the standardized effect sizes allowed by the alternative hypothesis. Consequently, Bayes factor functions (BFFs) may be viewed as mappings from standardized effect sizes to Bayes factors, or more formally as mappings from prior densities centered at standardized effect sizes to the corresponding Bayes factors \citep{Johnson2023}. Rather than reporting a single Bayes factor, BFFs summarize the continuum of Bayes factors obtained from a class of alternative priors and thus evaluate the evidence across a range of alternative hypotheses, eliminating the need to specify a particular value of the parameter under the alternative.}
{
\subsection{Primary Contribution}

We emphasize that the principal contribution of this work is the extension of the methodology proposed by \cite{Johnson2023} to the problem of { comparing null and alternative models corresponding to the presence or absence of partial correlation.} To the best of our knowledge, this is the first objective  Bayesian test proposed for { assessing evidence} regarding partial correlation coefficients that eliminates the need to specify subjective priors on nuisance parameters and does not require identifying variables as response or predictors, while still allowing prior knowledge about the parameter of interest to be incorporated by determining the prior mode based on the range of the parameter space under \(H_1\) and controlling for the dispersion around it.
 Although this extension constitutes a direct generalization of the existing approaches in \citep{Johnson2023, DATTA2025}, its implementation may not be immediately straightforward for practitioners. The primary difficulty arises because the previous literature consider residual based settings in which the sampling distribution of the test statistic under \(H_1\) follows a standard non-central distribution, such as the non-central \(t\) or \(F\) distribution. In contrast, when testing the significance of partial correlations in a regression-free setting, the distribution of the corresponding test statistic under the alternative hypothesis does not follow a non-central \(t\) or \(F\) distribution and is therefore non-standard. Consequently, a practitioner may inadvertently mis-specify the distribution of the test statistic under the alternative hypothesis. We also define the corresponding non-centrality parameter for this non-standard distribution, which quantifies the degree of departure from the null hypothesis; defining this parameter is itself a non-trivial task.  Furthermore, unlike \cite{Johnson2023, DATTA2025}, the resulting Bayes factor in this setting is not available in closed form, and its evaluation requires Monte Carlo–based computation of the marginal likelihood. To facilitate practical implementation, we provide code and functions that allow practitioners to compute the Bayes factor even when an analytical expression is unavailable. We emphasize that this provides a foundation for developing objective methodology for testing partial correlations in Gaussian Graphical Models (GGMs).

}

{ In Section \ref{secmethod}, we utilize these principles to construct BFFs for testing the presence of partial correlation. Section \ref{sec: prior_Comp} presents a comparison of alternative priors using the discrepancy measure of \cite{Wagenmakers2025}. We further evaluate our approach against existing Bayesian tests through an application to the \emph{Rapid Resumption dataset} \citep{LLERAS2011} in Section.}

\section{Methodology}\label{secmethod}

 In this section, we introduce an {objective Bayesian model comparison methodology between hypotheses corresponding to zero and non-zero partial correlations}. We further express the resulting Bayes factors based as a function of the standardized effect size $\omega$, defined as $\frac{\rho^*}{\sqrt{1-(\rho^*)^2}}$, where $\rho^*$ is the population partial correlation coefficient and $\omega$ is an increasing function of $\rho^*$. We compute a series of Bayes factors based on a specific test statistic for a sequence of alternatives. We emphasize that { assessing evidence for partial correlations} using Bayes factors constructed from test statistics is fundamentally different from a conventional \( F \)- or \( t \)-test. This distinction arises because the sampling distribution of the test statistic under the alternative hypothesis does not follow a standard parametric form. Furthermore, we explicitly define the non-centrality parameter of the test statistic under the alternative hypothesis and impose a nonlocal prior on this parameter to ensure principled Bayesian inference. 
 
 {The construction of BFFs proceeds in four steps: (i) define the data-generating model and partial correlation; (ii) identify a test statistic and derive its sampling distribution under both hypotheses ; (iii) place a non-local prior on the non-centrality parameter and compute the Bayes factor; (iv) express the Bayes factor as a function of the standardized effect size by linking the prior hyperparameters to $\omega$.}


{
\subsection{Model}\label{sec:model}

Let $\mathbf{X}_1, \mathbf{X}_2, \ldots, \mathbf{X}_n$ be a random sample of size $n$ from a $p$-variate normal distribution,
\begin{equation}\label{eq:model}
    \mathbf{X}_i \sim N_p(\boldsymbol{\mu}, \boldsymbol{\Sigma}), \qquad i = 1, 2, \ldots, n,
\end{equation}
where $\boldsymbol{\mu} \in \mathbb{R}^p$ is the population mean vector and $\boldsymbol{\Sigma}$ is the $p \times p$ positive definite population covariance matrix. Partition each observation as
\begin{equation}\label{eq:partition}
    \mathbf{X}_i = \begin{pmatrix} \mathbf{X}_{i1} \\ \mathbf{X}_{i2} \end{pmatrix},
\end{equation}
where $\mathbf{X}_{i1}$ is $q \times 1$ and $\mathbf{X}_{i2}$ is $(p - q) \times 1$. Correspondingly, partition the mean vector and covariance matrix as
\begin{equation}\label{eq:mu_partition}
    \boldsymbol{\mu} = \begin{pmatrix} \boldsymbol{\mu}_1 \\ \boldsymbol{\mu}_2 \end{pmatrix}, \qquad
    \boldsymbol{\Sigma} = \begin{pmatrix} \boldsymbol{\Sigma}_{11} & \boldsymbol{\Sigma}_{12} \\[4pt] \boldsymbol{\Sigma}_{21} & \boldsymbol{\Sigma}_{22} \end{pmatrix},
\end{equation}
where $\boldsymbol{\Sigma}_{11}$ is $q \times q$, $\boldsymbol{\Sigma}_{22}$ is $(p - q) \times (p - q)$, $\boldsymbol{\Sigma}_{12} = \boldsymbol{\Sigma}_{21}'$ is $q \times (p - q)$, and $\boldsymbol{\mu}_1 \in \mathbb{R}^q$, $\boldsymbol{\mu}_2 \in \mathbb{R}^{p-q}$. The sub-vector $\mathbf{X}_{i1}$ contains the variables whose partial correlation is of interest, and $\mathbf{X}_{i2}$ contains the $p - q$ variables to be conditioned upon.

\subsection{Population and Sample Partial Correlations}\label{sec:pc}

\subsubsection{Population partial correlation}\label{sec:pop_pc}

The conditional distribution of $\mathbf{X}_{i1}$ given $\mathbf{X}_{i2} = \mathbf{x}_2$ is 
\begin{equation}\label{eq:cond_dist}
    \mathbf{X}_{i1} \mid \mathbf{X}_{i2} = \mathbf{x}_2 \;\sim\; N_q\!\left(\boldsymbol{\mu}_1 + \boldsymbol{\Sigma}_{12}\boldsymbol{\Sigma}_{22}^{-1}(\mathbf{x}_2 - \boldsymbol{\mu}_2),\; \boldsymbol{\Sigma}_{11\cdot 2}\right),
\end{equation}
where the $q \times q$ conditional covariance matrix is $ \boldsymbol{\Sigma}_{11\cdot 2} \;=\; \boldsymbol{\Sigma}_{11} - \boldsymbol{\Sigma}_{12}\boldsymbol{\Sigma}_{22}^{-1}\boldsymbol{\Sigma}_{21}.$ Denote the $(s,t)$-th element of $\boldsymbol{\Sigma}_{11\cdot 2}$ by $\sigma_{st\cdot q+1,\ldots,p}$, for $s, t = 1, \ldots, q$. The \emph{population partial correlation} between the $s$-th and $t$-th components of $\mathbf{X}_{i1}$, holding $\mathbf{X}_{i2}$ fixed, is
\begin{equation}\label{eq:pop_pc}
    \rho^* \;\equiv\; \rho_{st\cdot q+1,\ldots,p} \;=\; \frac{\sigma_{st\cdot q+1,\ldots,p}}{\sqrt{\sigma_{ss\cdot q+1,\ldots,p}\;\sigma_{tt\cdot q+1,\ldots,p}}},
\end{equation}
for $s, t \in \{1, \ldots, q\}$, $s \neq t$. That is, $\rho^*$ is the ordinary Pearson correlation computed from the conditional covariance matrix $\boldsymbol{\Sigma}_{11\cdot 2}$.

\subsubsection{Sample partial correlation}\label{sec:sample_pc}

From the observed data, the total sum-of-products matrix is defined as
\begin{equation}\label{eq:A}
    A \;=\; \sum_{i=1}^{n}(\mathbf{X}_i - \bar{\mathbf{X}})(\mathbf{X}_i - \bar{\mathbf{X}})' \;=\; \begin{pmatrix} A_{11} & A_{12} \\ A_{21} & A_{22} \end{pmatrix},
\end{equation}
where $\bar{\mathbf{X}} = n^{-1}\sum_{i=1}^n \mathbf{X}_i$ is the sample mean, $A_{11}$ is $q \times q$, $A_{22}$ is $(p-q)\times(p-q)$, and $A_{12} = A_{21}'$ is $q \times (p-q)$. We define the $q \times q$ matrix
\begin{equation}\label{eq:A112}
    A_{11\cdot 2} \;=\; A_{11} - A_{12}A_{22}^{-1}A_{21},
\end{equation}
which is the Schur complement of $A_{22}$ in $A$ and serves as the sample analogue of $\boldsymbol{\Sigma}_{11\cdot 2}$. Denoting the $(s,t)$-th element of $A_{11\cdot 2}$ by $a_{st\cdot q+1,\ldots,p}$, the \emph{sample partial correlation} is defined as
\begin{equation}\label{eq:sample_pc}
    r^* \;\equiv\; r_{st\cdot q+1,\ldots,p} \;=\; \frac{a_{st\cdot q+1,\ldots,p}}{\sqrt{a_{ss\cdot q+1,\ldots,p}\;a_{tt\cdot q+1,\ldots,p}}},
\end{equation}
for $s, t \in \{1, \ldots, q\}$, $s \neq t$ (\cite{Anderson} Theorem~4.3.2). That is, the sample partial correlation is the normalized off-diagonal element of the $2 \times 2$ sub-matrix of $A_{11\cdot 2}$ corresponding to components $s$ and $t$, computed in exactly the same way that an ordinary correlation is computed from a covariance matrix.


\subsection{Definition and distribution of the test statistic}\label{sec:alt_r}
A widely adopted test statistic for assessing the significance of a partial correlation coefficient in the frequentist literature is given by:
\begin{equation}\label{eq:t_def}
    t_1 \;=\; r^*\,\sqrt{\frac{n - p + q - 2}{1 - r^{*2}}}.
\end{equation}
This is a monotone transformation of $r^*$ for $r^* \in (-1, 1)$. Under the null hypothesis $H_0\colon \rho^* = 0$,the statistic $t_1$ defined in \eqref{eq:t_def} follows, a central Student's $t$-distribution \citep{Anderson}:
\begin{equation}\label{eq:null_dist}
   t_1 \;\sim\; t_{n-p+q-2} \qquad \text{under } H_0\colon \rho^* = 0.
\end{equation}

Under the alternative hypothesis \( \rho^* \neq 0 \), the conditional density of \( t_1 \) is obtained via the transformation \( r^* = t_1 / \sqrt{t_1^2 + n - p + q - 2} \) applied to the conditional density of \( r^* \) given \( \rho^* \), yielding an expression of the form (by Theorem 4.3.5 of \cite{Anderson}),
\begin{align}\label{eq:density}
    f(r^* \mid \rho^*) &= \frac{(n-p+q-2)}{\sqrt{2\pi}}\,\frac{\Gamma(n-p+q-1)}{\Gamma\!\left(n-p+q-\tfrac{1}{2}\right)}\,(1-\rho^{*2})^{\frac{n-p+q-1}{2}}\,(1-r^{*2})^{\frac{n-p+q-4}{2}} \notag \\[6pt]
    &\quad \times\; (1 - \rho^*\,r^*)^{-(n-p+q-1) + \frac{1}{2}}\;{}_2F_1\!\left(\tfrac{1}{2},\,\tfrac{1}{2};\;n-p+q-\tfrac{1}{2};\;\tfrac{1 + \rho^*\,r^*}{2}\right),
\end{align}
for $-1 < r^* < 1$, where ${}_2F_1(a,b;\,c;\,x) = \sum_{j=0}^{\infty}\frac{\Gamma(a+j)\,\Gamma(b+j)\,\Gamma(c)}{\Gamma(a)\,\Gamma(b)\,\Gamma(c+j)}\,\frac{x^j}{j!}$
is the Gauss hypergeometric function. The density is valid provided $n \geq p - q + 4$. In the following lemma, we present the conditional density of \( t_1 \) given \( \rho^* \) under the alternative hypothesis.

\begin{lemma}\label{thm:t_density}
Under the model \eqref{eq:model} with $\rho^* \neq 0$, the density of $t_1$ defined in \eqref{eq:t_def} is
\begin{align}\label{eq:t_density}
    g(t_1 \mid \rho^*) &= \frac{\Gamma(n-p+q-1)}{\sqrt{2\pi}\;\Gamma\!\left(n-p+q-\tfrac{1}{2}\right)}\,\frac{(1 - \rho^{*2})^{\frac{n-p+q-1}{2}}\;(n-p+q-2)^{\frac{n-p+q}{2}}}{(t_1^2 + n-p+q-2)^{\frac{n-p+q-1}{2}}} \notag \\[6pt]
    &\quad \times\;\left(1 - \frac{\rho^*\,t_1}{\sqrt{t_1^2 + n-p+q-2}}\right)^{\!-(n-p+q-1) + \frac{1}{2}} \notag \\[6pt]
    &\quad \times\; {}_2F_1\!\left(\frac{1}{2},\,\frac{1}{2};\;n-p+q-\frac{1}{2};\;\frac{1}{2} + \frac{\rho^*\,t_1}{2\sqrt{t_1^2 + n-p+q-2}}\right),
\end{align}
for $t_1 \in (-\infty,\infty)$.
\end{lemma}

We parameterize the alternative through the non-centrality parameter \( \lambda = \sqrt{n - p + q - 2}\,\frac{\rho^*}{\sqrt{1 - \rho^{*2}}} \), which implies \( \rho^* = \frac{\lambda}{\sqrt{\lambda^2 + (n - p + q - 2)}} \). Substituting this relation, the conditional distribution of \(t_1\) given \(\lambda\) can be written as

\begin{align}\label{eqn:marginal_t1}
f(t_1 \mid \lambda) &= \frac{\Gamma(n-p+q-1)}{\sqrt{2\pi}\;\Gamma\!\left(n-p+q-\tfrac{1}{2}\right)}\,
\frac{(n-p+q-2)^{\,n-p+q-\frac{1}{2}}}{\left(\lambda^2+(n-p+q-2)\right)^{\frac{n-p+q-1}{2}}\,\left(t_1^2+(n-p+q-2)\right)^{\frac{n-p+q-1}{2}}} \notag \\[6pt]
&\quad \times\;\left(1 \;-\; \frac{\lambda\,t_1}{\sqrt{\left(\lambda^2+(n-p+q-2)\right)\left(t_1^2+(n-p+q-2)\right)}}\right)^{\!-(n-p+q-1)+\frac{1}{2}} \notag \\[6pt]
&\quad \times\;{}_2F_1\!\left(\frac{1}{2},\,\frac{1}{2};\;n-p+q-\frac{1}{2};\;\frac{1}{2} \;+\; \frac{\lambda\,t_1}{2\sqrt{\left(\lambda^2+(n-p+q-2)\right)\left(t_1^2+(n-p+q-2)\right)}}\right).
\end{align}
When $q=1$,
\begin{align}\label{eqn:marg_t1_q1}
f(t_1 \mid \lambda) &= \frac{\Gamma(n-p)}{\sqrt{2\pi}\;\Gamma\!\left(n-p+\tfrac{1}{2}\right)}\,
\frac{(n-p-1)^{\,n-p+\frac{1}{2}}}{\left(\lambda^2+(n-p-1)\right)^{\frac{n-p}{2}}\,\left(t_1^2+(n-p-1)\right)^{\frac{n-p}{2}}} \notag \\[6pt]
&\quad \times\;\left(1 \;-\; \frac{\lambda\,t_1}{\sqrt{\left(\lambda^2+(n-p-1)\right)\left(t_1^2+(n-p-1)\right)}}\right)^{\!-(n-p)+\frac{1}{2}} \notag \\[6pt]
&\quad \times\;{}_2F_1\!\left(\frac{1}{2},\,\frac{1}{2};\;n-p+\frac{1}{2};\;\frac{1}{2} \;+\; \frac{\lambda\,t_1}{2\sqrt{\left(\lambda^2+(n-p-1)\right)\left(t_1^2+(n-p-1)\right)}}\right).
\end{align}

Under these model specifications, lemma \ref{Two-sided_BF} gives the Bayes factor based on the test statistic, $t_1$ given a value of  hyperparameters $\left(\tau^2, \nu\right)$.

\begin{lemma}\label{Two-sided_BF}
Assume the distributions of a random variable \(t_1\) under the null and alternative hypotheses are described by
\begin{align}
    H_0: t_1 &\sim t_{n-p+q-2}(0), \\
    H_1: t_1 \mid \lambda &\sim f(t_1 \mid \lambda), \quad \lambda \mid \tau^2  \sim \pi_{nm}( \lambda\,|\, \tau^2, \nu),
\end{align}
where $\pi_{nm}( \lambda\,|\, \tau^2, \nu)$ denotes a normal moment prior as given by \eqref{eqn:mom_prior}. Then, the Bayes factor based on the test statistic, $t_1$, against the alternative is given by $BF_{10} = \frac{m_1(t_1 \mid \tau^2, \nu)}{m_0(t_1)}$, where,
\begin{eqnarray}
  m_1(t_1 \mid \tau^2, \nu) &=& \int_{-\infty}^{\infty} \frac{\Gamma(n-p+q-1)}{\sqrt{2\pi}\;\Gamma\!\left(n-p+q-\tfrac{1}{2}\right)}\,
\frac{(n-p+q-2)^{\,n-p+q-\frac{1}{2}}}{\left(\lambda^2+(n-p+q-2)\right)^{\frac{n-p+q-1}{2}}\,\left(t_1^2+(n-p+q-2)\right)^{\frac{n-p+q-1}{2}}} \notag \\[6pt]
&&\quad \times\;\left(1 \;-\; \frac{\lambda\,t_1}{\sqrt{\left(\lambda^2+(n-p+q-2)\right)\left(t_1^2+(n-p+q-2)\right)}}\right)^{\!-(n-p+q-1)+\frac{1}{2}} \notag \\[6pt]
&&\quad \times\;{}_2F_1\!\left(\frac{1}{2},\,\frac{1}{2};\;n-p+q-\frac{1}{2};\;\frac{1}{2} \;+\; \frac{\lambda\,t_1}{2\sqrt{\left(\lambda^2+(n-p+q-2)\right)\left(t_1^2+(n-p+q-2)\right)}}\right)\nonumber \\
  && \qquad \times \frac{|\lambda|^{2\nu}}{\left(2\tau^2\right)^{\nu/2} \Gamma\left(\nu + 0.5\right)} 
  \exp\left(-\frac{\lambda^2}{2\tau^2}\right) \ d\lambda,
\end{eqnarray}
and $m_0(.)$ denotes the density function of a central $t$-distribution with $n-p+q-2$ degrees of freedom.
\end{lemma}

We now express the Bayes factor in Lemma \ref{Two-sided_BF} as a function of the standardized effect size $\omega$, determined by the choice of the hyperparameter $\tau^2$.}

\subsection{Choice of \texorpdfstring{$\tau^2$}{tau-squared}}\label{sec:choose_tau}

Similar to the approach described in \cite{Johnson2023} and outlined in Section \ref{sec:BFF}, the parameter \(\tau^2\) is determined from the standardized effect size. The modes of the prior, as represented by the density in Equation \ref{nm_prior}, is given by \( \pm \sqrt{2\nu} \tau \). By equating the mode of the prior density to correspond to a function of standardized effect size ($\sqrt{n-p+q-2}\ \omega$), we have
\begin{equation}\label{BFF}
    \pm \sqrt{2\nu}\tau = \sqrt{n - p+q - 2} \, \omega, 
\end{equation}

which implies
\[
    \tau^2 = \frac{(n - p+q - 2) \  \omega^2}{2\nu}.
\]
Since the Bayes factor in Lemma \ref{Two-sided_BF} depends on the hyperparameters $(\tau^2, \nu)$, this formulation expresses the resulting Bayes factor as a function of the standardized effect size $\omega$.
\subsection{Choice of \texorpdfstring{$\nu$}{nu}}\label{subsec:choose_nu}
\begin{figure}[ht!]
    \centering
    \includegraphics[width=1\linewidth]{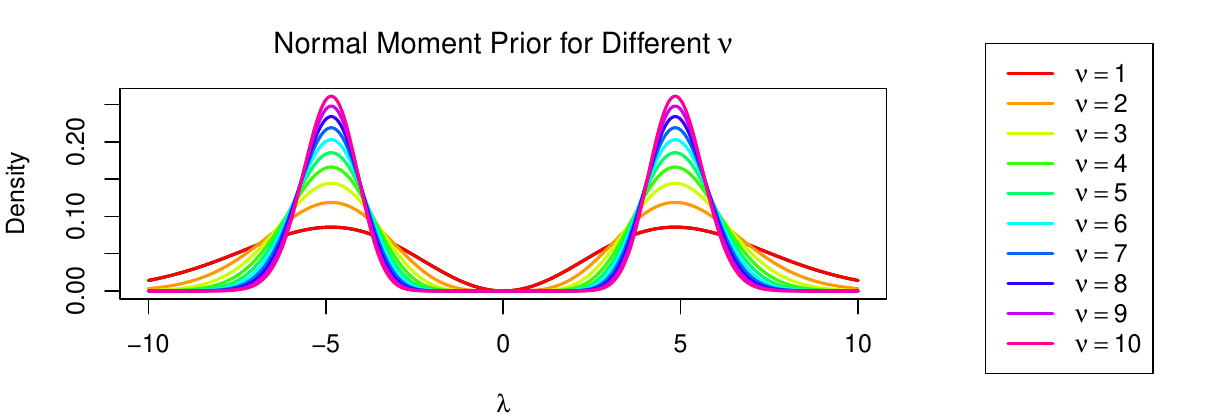}
    \caption{Normal Moment prior for various values of $\nu$.}
    \label{NM_prior}
\end{figure}

A method of moments (MOM) empirical Bayes estimator for \(\nu\) was proposed by \cite{DATTA2025} { for replicated experiments}. It is demonstrated that under the null hypothesis, the MOM estimator of \(\nu\) converges to 1. Additionally, they suggest that when informative subject-matter knowledge is available, selecting \(\nu > 1\) can help strengthen evidence in favor of the true hypothesis. However, in the absence of such prior knowledge or absence of replication, \(\nu = 1\) should be the default choice {since it ensures that the prior assigns sufficient mass across a wide range of plausible effect sizes.} Figure \ref{NM_prior} illustrates how \(\nu\) influences the variability around the prior mode.

{Since \(\nu\) controls the precision of the prior, that is, the variability around its mode, it should be chosen so that a large proportion of the prior mass is allocated to plausible effect sizes. \cite{Datta2025BFF} note that in many areas of the social and biological sciences prior knowledge regarding typical standardized effect sizes is often available. Investigators frequently focus on detecting effects within the standardized range \(0.2\) (small) to \(0.8\) (large), with \(0.5\) representing a medium effect size \citep{Cohen1988}. Accordingly, they recommend selecting \(\nu\) so that the alternative prior on the noncentrality parameter \(\lambda\) places a specified proportion of its mass—say \(\gamma \approx 0.9\)—within the range corresponding to standardized effects in \((0.2,0.8)\), assuming that the prior mode for \(\omega\) is \(0.5\). \cite{Datta2025BFF} discuss the choice of the dispersion hyperparameter \(\nu\) in the context of inverse moment priors. We follow the same principle in determining the prior dispersion around the mode. Specifically, for a given effect size \(\omega\), \(\nu\) is chosen so that \(P(\lambda \in [a,b])=\gamma=0.9\), where the range \([a,b]\), or  the value \(\gamma\), is determined based on available prior knowledge. In the absence of such prior information, we recommend the default choice \(\nu=1\) since that assigns positive mass on a wide range of effect sizes. A comparison of the class of alternative priors for $\nu = 1,5,10$ is presented in the following section. } 

In our real-data application, we observed that any value of \(\nu \geq 1\) allocates over 95\% of the prior mass to effect sizes in the interval $ \omega \pm 0.3$, where $\omega$ is any standardized effect size that determines the prior mode. Therefore, we set \(\nu = 1\) in our real data.

{
\section{Comparison of alternative class of priors}\label{sec: prior_Comp}

The asymptotic behavior of Bayes factors under a true point null hypothesis is determined by the behavior of the point alternative prior density in a neighborhood of the null value of the parameter being tested \citep{Johnson2010, Datta2025BFF}. In the present setting, we consider testing whether the non-centrality parameter $\lambda$ of the test statistic equals zero; consequently, the local behavior of the prior distribution for $\lambda$ near zero plays a central role in governing the properties of the resulting Bayes factor. \cite{Johnson2010, DATTA2025} show that, under suitable regularity conditions, the Bayes factor in favor of the alternative hypothesis obtained using a Normal moment prior exhibits polynomial decay when the null hypothesis is true. Specifically, for the \(z\) and \(t\) tests the Bayes factor against the null hypothesis satisfies \(BF_{10}=O_p(n^{-\nu-1/2})\), while for the \(\chi^2_k\) and \(F_{k,m}\) tests it satisfies \(BF_{10}=O_p(n^{-\nu-k/2})\) \citep{DATTA2025}. When the alternative hypothesis is true, the Bayes factor against the alternative hypothesis typically satisfies \(BF_{01}=O_p(\exp(-cn))\) for some \(c>0\), provided the prior under the alternative is continuous and positive at the true non-null parameter value \citep{Bahadur}. The rates for the \(z\) and \(t\) tests also hold in the present framework since the same regularity conditions are applicable. 

In finite samples, asymptotic convergence rates may be misleading, and practical performance may differ substantially from asymptotic expectations. Under true alternative hypotheses, $BF_{01}$ is typically $O_p(\exp(-cn))$ for some $c>0$ for all alternative priors that are continuous and positive at the true non-null parameter value \citep{Bahadur}.

To assess the finite-sample performance of Bayes factors, we draw on a recent result of \citet{Wagenmakers2025}, who established the following identity. Let $f$ and $g$ denote the prior predictive distributions under two competing models, ${\cal M}_f$ and ${\cal M}_g$, each assigned equal prior probability. They define the discrepancy measure
\begin{equation*}
\DEP = \int f(x)\frac{f(x)}{f(x)+g(x)}\,dx = \int g(x)\frac{g(x)}{f(x)+g(x)}\,dx,
\end{equation*}
with $\DEP = \PED$. When equal prior probability is assigned to the two models, this identity implies that the expected posterior probability of ${\cal M}_f$ when ${\cal M}_f$ is true equals the expected posterior probability of ${\cal M}_g$ when ${\cal M}_g$ is true. We write $\dep$ as shorthand for $\DEP$ when the compared models are clear from context. This measure can be used to compare BFFs induced by different alternative prior distributions. Specifically, letting ${\cal M}_0$ denote the null model $H_0$, and ${\cal M}_{1a}$ and ${\cal M}_{1b}$ denote two alternative models, we compare ${\rm D}_{EP}(0\parallel 1a)$ and ${\rm D}_{EP}(0\parallel 1b)$. If ${\rm D}_{EP}(0\parallel 1a) > {\rm D}_{EP}(0\parallel 1b)$, then the expected posterior probability of the true model when the null is compared to ${\cal M}_{1a}$ is higher than the expected posterior probability of the true model when the null is compared to ${\cal M}_{1b}$, assuming equal prior probabilities are assigned to each model in both comparisons. $\dep$ takes values in the interval $(0.5, 1)$. Values of $\dep$ close to 1 indicate that the Bayes factor exhibits strong discriminatory capability, thereby yielding greater expected support for the true hypothesis. In contrast, values of $\dep$ near 0.5 suggest that, under the given prior specification, the data provide limited information for distinguishing between the competing models and hence the Bayes factor is uninformative on an average.

A convenient way to parameterize non-local priors, when studying their efficiency in accumulating evidence against a null hypothesis, is through their modes, which serve as a natural indexing quantity. This approach, however, is not applicable to local alternative priors commonly employed in standard Bayesian null hypothesis testing, as these are typically centered at the null value, rendering the mode uninformative for indexing. Moreover, some prior distributions lack finite moments, precluding parameterization via means or variances. To address these limitations, one may instead index alternative priors using their interquartile range (IQR) in two-sided settings and their median in one-sided settings. Such a parameterization yields a consistent and interpretable framework for comparing the performance of BFFs across diverse classes of prior distributions. We consider three alternative prior specifications for assessing the presence or absence of a partial correlation as follows:

\begin{itemize}
    \item[1.] Normal moment prior for $\nu = 1,5,10$.
    \item[2. ] Effect size priors \citep{KlauerMeyerGrantKellen2025}: The effect-size prior imposed on the non-centrality parameter $\lambda$ is
\begin{align}\label{eqn:esprior}
    \pi_{es}(\lambda \mid \nu_k,\lambda_e,s) =
\frac{1}{2}\,f_t(\lambda \mid \nu_k,-\lambda_e,s) + \frac{1}{2}\,f_t(\lambda \mid \nu_k,\lambda_e,s),
\end{align}

where $f_t(\lambda \mid \nu_k,\mu,s)$ denotes the density of a shifted, scaled $t$ distribution with $\nu_k$ degrees of freedom, location parameter $\mu$, and scale parameter $s$, that is, $
f_t(\lambda \mid \nu_k,\mu,s) =\frac{1}{s}\,dt\left(\frac{\lambda-\mu}{s}, df = \nu_k\right),
$ with
$\lambda_e=\sqrt{df}\,\omega_e,
s=\sqrt{\frac{\nu_k-2}{\nu_k}}\,\lambda_e.$

\item[3.] Stretched-$\beta$ prior \citep{Kucharsky2023PartialCorrelation}: We consider a stretched beta prior on the population partial correlation $\rho^* \in (-1,1)$ of the form
\begin{align}\label{eqn: sbprior}
\pi_{SB}(\rho^* \mid \alpha)
=
\frac{(1-(\rho^*)^2)^{\alpha-1}}{2^{\,2\alpha-1} B(\alpha,\alpha)},
\end{align}
where $\alpha > 0$ controls the concentration of mass around zero. Since noncentrality parameter is defined as $\lambda = \sqrt{df}\,\rho^*/\sqrt{1-(\rho^*)^2}$, the corresponding prior on $\lambda \in \mathbb{R}$ is given by
\begin{align}
\pi_{SB}(\lambda \mid \alpha, df)
=
\frac{df^{\alpha}}{2^{\,2\alpha-1} B(\alpha,\alpha)}
(\lambda^2 + df)^{-(\alpha + 1/2)}.
\end{align} 
\end{itemize}

The $\dep$ curves corresponding to these alternative prior specifications, relative to the null hypothesis, are displayed in Fig.~\ref{fig:DEP}. For the Stretched-$\beta$ $(\alpha,\alpha)$ prior, symmetry implies that $Q_{0.25}(\lambda) = -Q_{0.75}(\lambda)$, and hence the interquartile range satisfies $\mathrm{IQR}_\lambda(\alpha) = 2Q_{0.75}(\lambda)$. Under the induced prior on $\rho^*$, the $0.75$ quantile is given by $Q_{0.75}(\rho^*) = 2F^{-1}_{\mathrm{S\text{-}Beta}(\alpha,\alpha)}(0.75) - 1$, where $F^{-1}_{\mathrm{S\text{-}Beta}(\alpha,\alpha)}$ denotes the quantile function of the $\mathrm{Stretched}\text{-}\beta(\alpha,\alpha)$ distribution. Using the transformation $\lambda = \sqrt{df}\,\rho^*/\sqrt{1-(\rho^*)^2}$, it follows that $Q_{0.75}(\lambda) = \sqrt{df}\,\dfrac{Q_{0.75}(\rho^*)}{\sqrt{1 - Q_{0.75}(\rho^*)^2}}$. Consequently, the interquartile range on the $\lambda$ scale is given by $\mathrm{IQR}_\lambda(\alpha) = 2\sqrt{df}\,\dfrac{2F^{-1}_{\mathrm{Beta}(\alpha,\alpha)}(0.75) - 1}{\sqrt{1 - \left(2F^{-1}_{\mathrm{Beta}(\alpha,\alpha)}(0.75) - 1\right)^2}}$. The values of $\tau$ corresponding to a specified IQR for the normal moment priors were obtained numerically. For the effect size priors, we fix $\nu_k = 3$ and calibrate $\omega_e$ to match the desired IQR.

\begin{figure}[ht!]
    \centering
    \includegraphics[width=0.8\linewidth]{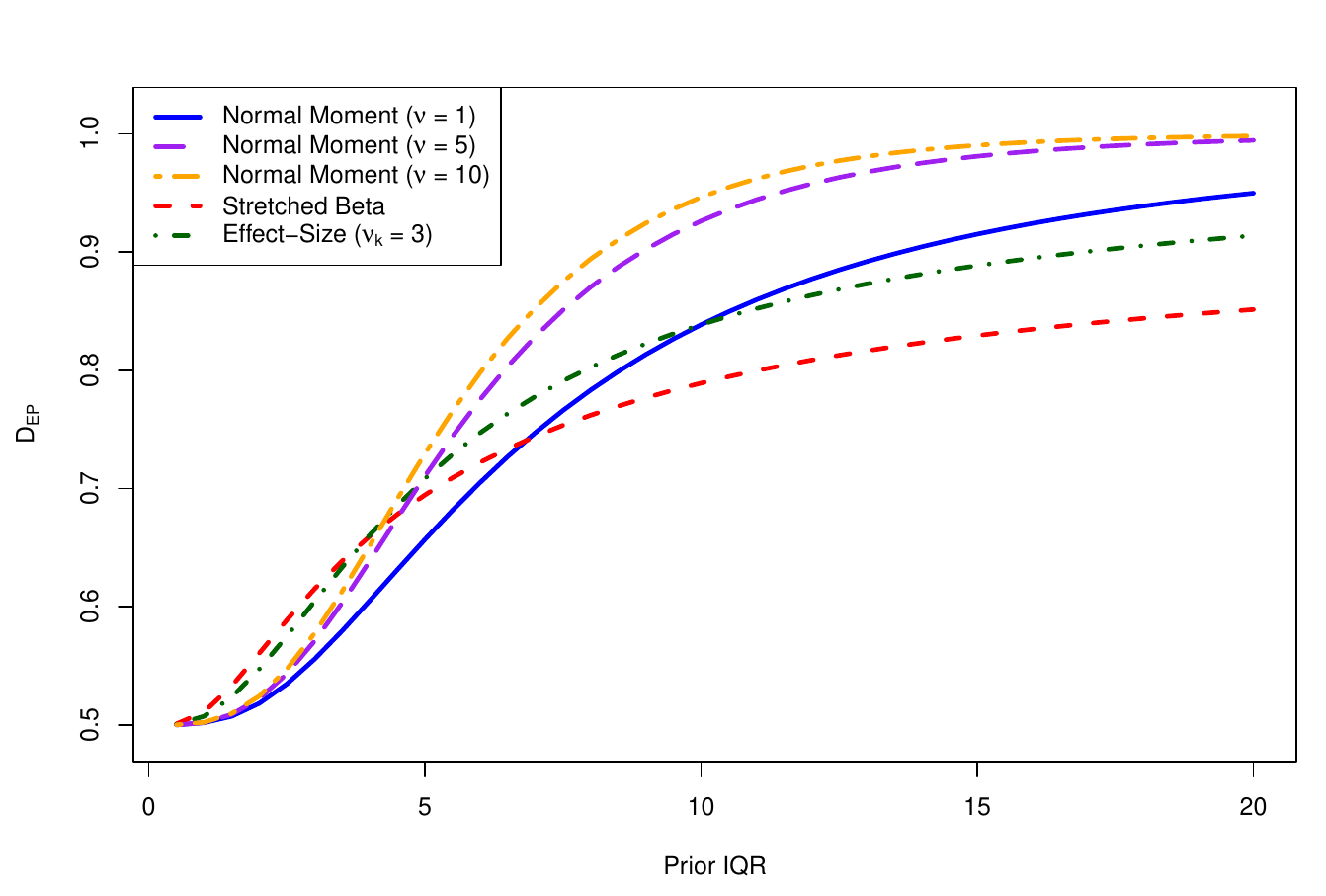}
    \caption{$\dep$ plots for $t_1$-test statistic with 38 degrees-of-freedom. The horizontal axis provides the IQR that each prior density assigns around the null value of the non-centrality parameter (i.e., $\lambda=0$).}
    \label{fig:DEP}
\end{figure}

Fig.~\ref{fig:DEP} illustrates several contrasts among the $\dep$ values associated with the three alternative prior specifications. When the IQR of the priors exceeds approximately $5.0$, the normal moment priors of orders $5$ and $10$ yield substantially larger $\dep$ than the local priors. Noting that the noncentrality parameter of the statistic $t_1$ is $\lambda = \sqrt{n-p-1}\,\omega$, an IQR of $5$ corresponds, for example, to a sample size of $40$, $p=1$, and a standardized effect size of $\omega \approx 0.4$--$0.5$, or equivalently $\rho^* \approx 0.3$--$0.4$, representing a medium standardized effect \citep{Cohen1988}. For smaller values of the prior IQR (IQR $< 4$), the Stretched-$\beta$ and effect size priors exhibit slightly improved performance relative to the normal moment priors, although none of the priors provide strong evidence in favor of the true model. For larger values of the IQR (IQR $> 10$), all normal moment priors yield greater evidence supporting the true model compared to the local priors. In the intermediate regime, with IQR between $5$ and $10$, the normal moment priors with $\nu = 5, 10$ outperform the remaining prior specifications. Overall, as the IQR of the prior distributions increases, the normal moment prior shifts probability mass away from the origin at a faster rate than the corresponding local priors. This behavior leads to a more rapid increase in the ${\rm D_{EP}}$ statistic for the normal moment priors, as illustrated in Fig.~\ref{fig:explanation}. This trend continues as the IQR increases further, with the normal moment priors providing progressively stronger support for the true hypothesis.

\begin{figure}[ht!]
    \centering
    \includegraphics[width=1\linewidth]{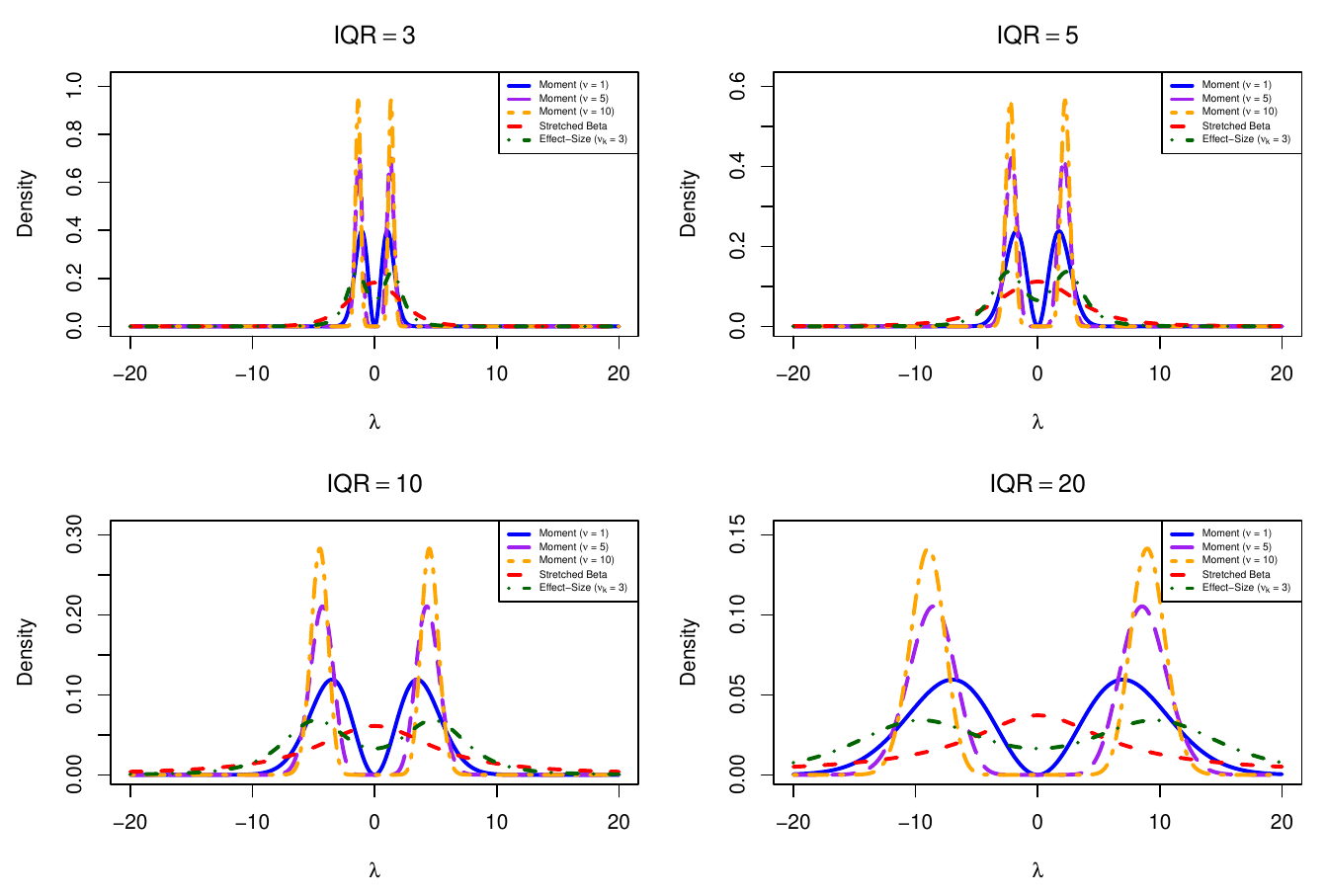}
    \caption{Plots of prior densities:  The IQRs of the normal moment, effect size and stretched-$\beta$ priors range from 3 (upper left) to 20 (lower right). }
    \label{fig:explanation}
\end{figure}
 }

\section{Application: The rapid resumption data}\label{sec:application}

In a study by Lleras, Porporino, Burack, and Enns \citep{LLERAS2011}, the role of implicit prediction in visual search was investigated using an interrupted search task across different age groups (7, 9, 11, and 19 years old). In the experiment, participants engaged in a conventional interrupted search task, where they had brief 500-ms glimpses of a display screen, interspersed with 1000-ms periods of a blank screen. During each glimpse, the display contained 15 "L" shapes (as distractors) and one "T" shape (as the target). The shapes were evenly split in color between red and blue, with the target "T" randomly assigned one of these colors. The objective for the participants was to quickly identify and report the color of the "T" shape by pressing one of two designated keys. The study aimed to understand the rapid resumption phenomenon, where subsequent looks at the stimulus within 500 ms significantly increase correct response rates, compared to the initial look. Analyzing the observations, correlation ($r_{XY} = 0.51, p < 0.01$) was indicated between the average successful search time($X$) and the rate of rapid resumption responses($Y$). However, recognizing the potential influence of age($Z$) on these variables (with high correlations between search time and age, $ r_{XZ} = -0.78$, and rapid resumption and age, $r_{YZ} = -0.66$), the researchers calculated a partial correlation to control for the age effect, which turned out to be $r_{XY \mid Z} = -0.01$. The corresponding $t$ statistic ($t_1$) was $-0.06$ on 37 degrees of freedom, which resulted in a $p-$value of 0.95. This finding left the null-hypothesis unrefuted.

{ In this section, we compare the evidence obtained under the normal moment prior (MoM prior) with $\nu = 1$, the effect size prior with $\nu_k = 3$, each specified on the non-centrality parameter $\lambda = \sqrt{n-p-1}\,\omega$ and the Stretched-$\beta$ prior with $\alpha = 0.5$ applied on $\rho^*$. The Bayes factor under the Stretched-$\beta$ prior is computed from the full data likelihood and does not rely on a test statistic. For the normal moment and effect size priors, the prior modes are varied over a range of $\omega \in (-1.33, 1.33)$. Corresponding to this sequence of standardized effect sizes, we determine a sequence of values of $\tau$ for the MoM prior using the procedure described in Section \ref{sec:choose_tau}, along with a range of values of $\lambda_e$ in \eqref{eqn:esprior}, representing the modes of the effect size prior, obtained via $\lambda_e = \sqrt{n-p-1}\,\omega$. This construction yields Bayes factor functions as functions of the effect size, indexed by alternative specifications with prior modes centered at standardized effect sizes. In contrast, since the Stretched-$\beta$ prior has its mode fixed at $0$, Bayes factor functions cannot be obtained, and only a single Bayes factor against the null hypothesis is available.  } 



\begin{figure}
    \centering
    \includegraphics[width=0.8\linewidth]{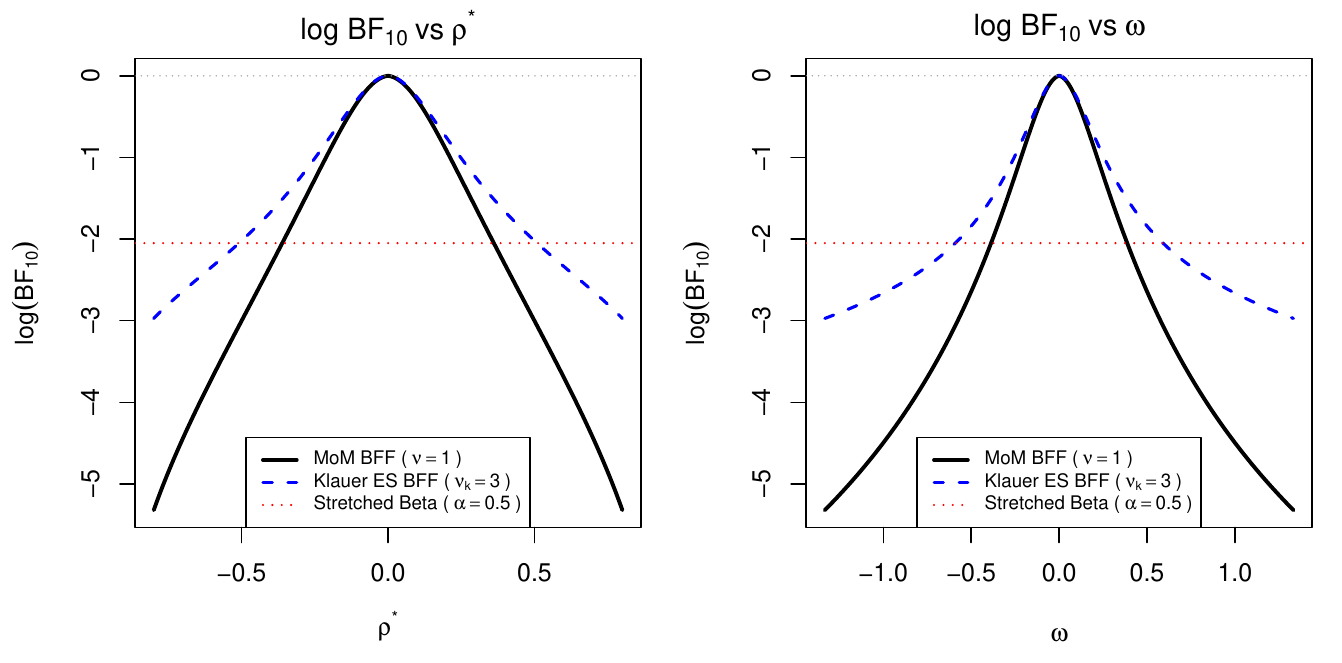}
    \caption{Logarithm of the Bayes factor function in favor of the alternative hypothesis plotted against (a) $\rho^*$  and (b) $\omega = \frac{\rho^*}{\sqrt{1-(\rho^*)^2}}$ using first order normal moment prior (black), effect size prior (blue dashed) and stretched-$\beta$ prior with $\alpha = 0.5$ (red dotted) on the non-centrality parameter.}
    \label{BFF_PCC}
\end{figure}

{Figure \ref{BFF_PCC} displays the Bayes factor based on the test statistic $t_1$, plotted as a function of $\omega$ and $\rho^*$ under the MoM prior and the effect size prior. The Bayes factor corresponding to the Stretched-$\beta$ prior applied to $\rho^*$ using the full data likelihood is additionally indicated by a horizontal red line}. 

{ For $|\rho^*| < 0.2$ or $|\omega| < 0.2$ (noting that $\rho^*$ and $\omega$ are approximately comparable for small effects), the logarithm of the Bayes factor against the null hypothesis under the MoM prior lies in the range $(0, -0.83)$, whereas under the effect size prior it lies in $(0, -0.081)$. This indicates that the MoM prior yields stronger evidence in favor of the null hypothesis even for small effect sizes. For moderate to larger effect sizes, the Bayes factor based on the MoM prior continues to favor the null hypothesis more strongly relative to the other two prior specifications. The logarithm of the Bayes factor against the null obtained under the stretched-$\beta$ prior is $-2.04$. For small effect sizes, the stretched-$\beta$ prior yields comparatively stronger evidence in favor of the null hypothesis. In contrast, for moderate to large effect sizes, both the MoM prior and the effect size priors accumulate greater evidence supporting the null.}

{
\section{Extension to Gaussian Graphical Models}

The methodology developed in this article for { assessing evidence} of a single partial correlation coefficient can, in principle, be extended to Gaussian graphical models (GGMs), although such an extension is not immediate. The reason is that the present procedure addresses inference on a single parameter $\rho^\ast$, whereas Gaussian graphical model estimation involves simultaneous inference on many partial correlations and, more importantly, uncertainty regarding the entire graph structure. Nevertheless, the connection between the present methodology and GGMs is direct because edges in a Gaussian graphical model correspond precisely to non-zero partial correlations.

Consider multivariate observations $\mathbf{Z}_1,\ldots,\mathbf{Z}_n$ sampled independently from $\mathcal{N}_p(\boldsymbol{\mu},\boldsymbol{\Sigma})$. Let $\boldsymbol{\Omega}=\boldsymbol{\Sigma}^{-1}$ denote the precision matrix. In a Gaussian graphical model, the conditional independence structure of the variables is determined by the entries of $\boldsymbol{\Omega}$. In particular, two variables $i$ and $j$ are conditionally independent given the remaining variables if and only if $\Omega_{ij}=0$. Equivalently, the partial correlation between variables $i$ and $j$ conditional on all remaining variables satisfies $\rho_{ij\mid -ij}=0$. Consequently, the presence of an edge between nodes $i$ and $j$ in the graph is equivalent to the hypothesis that the corresponding partial correlation coefficient differs from zero. Therefore, estimation of a Gaussian graphical model may be interpreted as simultaneously testing a collection of hypotheses of the form $H_0:\rho_{ij\mid -ij}=0$ for all $1\le i<j\le p$.

Under the multivariate normal model, the sample partial correlation coefficient between variables $i$ and $j$ conditional on the remaining variables is denoted by $r^\ast_{ij}$. The standard frequentist test statistic for testing $H_0:\rho_{ij\mid -ij}=0$ is
\begin{align}
t_{ij}=\frac{\sqrt{n-p-1}\,r^\ast_{ij}}{\sqrt{1-r_{ij}^{\ast 2}}},
\end{align}
which follows a central $t$ distribution with $n-p-1$ degrees of freedom under the null hypothesis. The methodology developed in this paper can therefore be applied to each pair $(i,j)$ by computing the Bayes factor function associated with this statistic. Specifically, if $\omega_{ij}$ denotes the standardized effect size corresponding to the population partial correlation $\rho_{ij\mid -ij}$, then the Bayes factor function $BF_{ij}(\omega_{ij})$ provides a measure of evidence against the null hypothesis across a range of alternative effect sizes. In this way, the methodology yields a collection of Bayes factors, one for each pair of variables, quantifying the evidence for conditional dependence between the corresponding nodes.

A simple conceptual extension to Gaussian graphical model estimation can therefore be described as follows. For each pair of variables $(i,j)$, one first computes the sample partial correlation coefficient $r^\ast_{ij}$ and the corresponding test statistic $t_{ij}$. Using the framework developed in this article, a Bayes factor function $BF_{ij}(\omega_{ij})$ is then calculated. Evidence for conditional dependence between variables $i$ and $j$ can then be summarized by the magnitude of this Bayes factor. The adjacency matrix of the graphical model could be obtained by declaring an edge between nodes $i$ and $j$ whenever the evidence in favor of the alternative hypothesis exceeds a predetermined threshold. In other words, if $BF_{ij}$ denotes the Bayes factor in favor of $H_1:\rho_{ij\mid -ij}\neq0$, the graph could be defined through the rule
\begin{align}
\text{Edge}(i,j)=
\begin{cases}
1 & \text{if } BF_{ij}>c,\\
0 & \text{otherwise},
\end{cases}
\end{align}
for some evidence threshold $c$. Applying this procedure to all $\frac{p(p-1)}{2}$ pairs of variables yields an estimated adjacency matrix describing the conditional dependence structure of the variables.

Although the above construction illustrates how the proposed Bayes factor functions can be used to evaluate conditional associations in a network, an important complication arises when moving from inference on a single partial correlation coefficient to inference on an entire graphical model. In particular, a Gaussian graphical model involves uncertainty over the entire graph rather than over individual edges. A graph with $p$ nodes contains $\frac{p(p-1)}{2}$ potential edges and therefore $2^{p(p-1)/2}$ possible graph structures. The pairwise testing procedure described above treats each edge independently, whereas in reality the partial correlations $\rho_{ij\mid -ij}$ depend on the full covariance structure of the variables. Consequently, the presence or absence of edges in a graphical model is not independent across pairs of variables.

A second difficulty concerns multiplicity. Because the number of potential edges grows quadratically with the number of variables, performing independent tests for each edge may lead to incoherent inference at the level of the entire graph. In standard Bayesian formulations of Gaussian graphical models, this issue is addressed by defining a likelihood for the precision matrix conditional on a graph structure and placing prior distributions on both the graph and the precision matrix. For example, Bayesian GGM methods often employ $G$-Wishart priors or spike-and-slab priors on the entries of the precision matrix, and inference is performed by integrating over both graph structures and precision matrices. In contrast, the present methodology operates directly on the sampling distribution of a test statistic and therefore does not explicitly incorporate a prior distribution over graphs.

For this reason, a principled extension of the proposed Bayes factor function methodology to Gaussian graphical models would require combining the edge-specific Bayes factors with a prior distribution over graph structures. One possible formulation is to introduce a binary indicator $G_{ij}$ representing the presence of an edge between nodes $i$ and $j$. If $\pi_{ij}= P(G_{ij}=1)$ denotes a prior probability for the inclusion of the edge, the posterior probability of an edge could be defined through a relationship of the form $P(G_{ij}=1\mid\text{data})=\frac{BF_{ij}\pi_{ij}}{BF_{ij}\pi_{ij} + (1-\pi_{ij})} $. Additional structure could then be imposed through a prior distribution on the entire adjacency matrix, such as a sparse Bernoulli prior encouraging graphs with relatively few edges. The resulting posterior distribution over graphs could then be summarized using posterior inclusion probabilities or by selecting a maximum a posteriori graph.

Despite these challenges, the Bayes factor function methodology developed here offers two advantages that make it attractive for graphical model applications. First, because the approach is based on the sampling distribution of test statistics rather than on the full likelihood, it avoids the need to specify prior distributions on nuisance parameters such as covariance or precision matrices. Second, the computation of Bayes factors based on test statistics is straightforward and therefore substantially simpler than the Markov chain Monte Carlo procedures typically required for Bayesian graphical model estimation. As a result, the proposed methodology could potentially be used as a computationally efficient screening tool for identifying candidate edges in large graphical models.

Developing a coherent framework that integrates Bayes factor functions with prior distributions over graph structures represents a promising direction for future research. Such an extension would allow the objective Bayesian testing procedure proposed in this article to be used not only for testing individual partial correlations but also for learning the conditional dependence structure of multivariate systems represented by Gaussian graphical models.}
\section{Discussion}

{In this article, we present an objective Bayesian test for partial correlation and evaluate its effectiveness by comparing it to existing methods through a discrepancy measure($\dep$) and real data analysis. Our primary contribution is the formulation of a Bayes factor based on a residual-free summary test statistics for partial correlations, eliminating the need to specify subjective priors on nuisance parameters, which can influence inference when Bayes factors are derived using the full dataset. This approach also serves as an interesting application of Bayes factor functions (BFF) \citep{Johnson2023, DATTA2025}, enabling the computation of posterior odds\eqref{eq: BF-postodds} across a sequence of alternative hypotheses. It is important to note that the proposed test is not a conventional ($t/F$) test, as the distribution of the test statistic under the alternative hypothesis follows a non-standard form.}

{We formulate a framework for defining Bayes factor functions to test partial correlations, wherein the Bayes factor against the null hypothesis is expressed across a sequence of alternative hypotheses. In doing so, we provide an objective specification of the prior hyperparameters. In particular, the hyperparameter $\tau^2$ is specified across a range of plausible alternatives, thereby eliminating the need to specify a simple alternative hypothesis in order to determine the prior distribution. }

{ A comparison of the MoM prior with the effect size and Stretched-$\beta$ priors indicates that, for moderate to large prior IQR values around the null, the MoM prior accumulates evidence at a faster rate than the other two. This is consistent with the fact that, under the true null hypothesis, Bayes factors based on local priors converge at a rate of $\sqrt{n}$ \citep{Johnson2010, Pramanik2024}, whereas those based on MoM priors converge at a rate of $n^{-\nu-\frac{1}{2}}$\citep{DATTA2025}. }

We begin by defining a Bayes factor based on the test statistic. When observations are sampled from a multivariate normal distribution, the frequentist test statistic \( t_1 \) follows a central \( t \)-distribution under the null hypothesis, regardless of the sample size. Under the alternative hypothesis, we derive the density of \( t_1 \), where the non-centrality parameter is given by \( \lambda = \sqrt{n - p - 1} \frac{\rho^*}{\sqrt{1 - (\rho^*)^2}} \). We impose a normal moment prior on $\lambda$\citep{Johnson2010} where the scale parameter is defined through $\tau$. $\tau$ is chosen objectively by equating the prior mode to a function of the standardized effect size(deterministic quantity), $\omega = \frac{\rho^*}{\sqrt{1 - (\rho^*)^2}}$\citep{Johnson2023}. One could argue that defining our Bayes factors based on a summary test statistic leads to some loss of information. However, this is balanced by the necessity of specifying a subjective prior, which inherently involves making subjective choices regarding the prior hyperparameters \citep{Johnson2005}.  {  This observation is further substantiated by the real data application in Section \ref{sec:application}, where for medium to large effect sizes, both BFBOTS obtained using MoM and effect size priors accumulate greater evidence against the alternative hypothesis compared to the full-data based Bayes factor obtained using the Stretched-$\beta$ prior.}

{ It is worth noting that when the observations do not arise from a Gaussian distribution, for large sample sizes, applying Fisher’s transformation to the sample partial correlation coefficient $r^*$,
\[
z = \frac{1}{2}\log\!\left(\frac{1 + r^*}{1 - r^*}\right),
\]
facilitates the use of a Bayes factor based on a $z$-test statistic \citep{Johnson2023}. Under $H_0$, we have $z \underset{\text{asymp}}{\sim} N\!\left(0, \frac{1}{\sqrt{n-p+q-2}}\right)$, whereas under $H_1$, $z \underset{\text{asymp}}{\sim} N\!\left(\frac{1}{2}\log\!\left(\frac{1 + \rho^*}{1 - \rho^*}\right), \frac{1}{\sqrt{n-p+q-2}}\right)$. Consequently, under the alternative hypothesis, the mean (non-centrality parameter) of the test statistic is given by $\omega = \frac{1}{2}\log\!\left(\frac{1 + \rho^*}{1 - \rho^*}\right)$. It follows that $\sqrt{n-p+q-2}\,z \underset{\text{approx}}{\sim} N\!\left(\frac{\sqrt{n-p+q-2}}{2}\log\!\left(\frac{1 + \rho^*}{1 - \rho^*}\right), 1\right)$ under $H_1$, and a normal moment prior may be imposed on $\lambda = \frac{\sqrt{n-3}}{2}\log\!\left(\frac{1 + \rho^*}{1 - \rho^*}\right)$ to construct a Bayes factor function based on the $z$-test statistic, as described in \cite{Johnson2023, DATTA2025}.}

Similar comparisons were made by \cite{Pramanik2024}, where the Jeffrey-Zellner-Siow (JZS) prior was used as a competing method against Non-local alternative prior densities. Their findings demonstrated that the JZS prior accumulates evidence in favor of the true null hypothesis at a rate of only \(\sqrt{n}\). In contrast, as shown in \citet{DATTA2025}, moment priors accumulate evidence in favor of the true null hypothesis at a rate of \(n^{\nu + \frac{1}{2}}\). { The proposed methodology is applicable when the observations arise from a multivariate normal distribution, or when the sample size is sufficiently large so that a Fisher's \(z\) transformation can be applied. At the same time, this framework provides a foundation for developing objective Bayesian tests in more general settings in future work. }

Overall, we propose an objective Bayesian test for partial correlation coefficients that accumulates evidence in favor of the true alternative hypothesis at an exponential rate and at polynomial rate in favor of the true null hypothesis. Future research directions include utilizing Bayes Factor functions (BFFs) for testing partial correlations in a nonparametric setting and extending our methodology to  Graphical Models. {The code used to implement the experiments demonstrated in Sections \ref{sec: prior_Comp} and \ref{sec:application} can be accessed through \url{https://github.com/Saptati-Datta/BFF_PCC}.}

\section{Acknowledgment}
I am deeply grateful to my supervisor, Dr. Valen E. Johnson for providing invaluable insights that greatly contributed to the development of this manuscript.

\section{Funding}
I acknowledge support from NSF Grant DMS-2311005. 
\bibliographystyle{apalike}
\bibliography{cas-refs}

\end{document}